# Time series analysis of Parkinson's disease, Huntington's disease and Amyotrophic Lateral Sclerosis

Farshad Merrikh-Bayat[a*]

[a]*Department of Electrical and Computer Engineering, Zanjan University, Zanjan, Iran*

**Abstract**

The aim of this paper is to study the (clinical) time-series data of three diseases with complex dynamics: Parkinson's disease, Huntington's disease and Amyotrophic Lateral Sclerosis. For this purpose, first all of the time series data are embedded in a vector space of suitable dimension and then the correlation dimension of the above mentioned diseases is estimated. The results are also compared with healthy control subjects. At the next step, existence of chaos in these diseases is investigated by means of the so-called 0-1 test. The simulations show that none of the above mentioned diseases are chaotic.

## 1. Introduction

The aim of this paper is to analyze the time series data of three diseases with complex dynamics- Parkinson's disease, Huntington's disease and Amyotrophic Lateral Sclerosis (ALS). Clearly, the results obtained from these analyses may be helpful for better understanding the pathophysiology of these three diseases and to improve our ability to measure responses to therapeutic interventions. The reader is referred to [1] and [2] for more information about the dynamics of Huntington's disease and ALS, respectively.

In the following, first we will estimate the correlation dimension of the above mentioned diseases and then apply the so-called 0-1 test to investigate the chaotic behavior in these three diseases.

---

* Corresponding author. Tel.: +98-241-5154061.
 *E-mail address*: f.bayat@znu.ac.ir.

## 2. Clinical Findings

All the time-series data used in this paper are downloaded from [3]. The raw data were obtained using force-sensitive resistors, with the output roughly proportional to the force under the foot. Ten records for Parkinson's disease, ten for Huntington's disease, ten for ALS, and ten for healthy control subjects are employed in this paper. Each of these records contains approximately 90'000 samples with the sampling period of 0.0035 s. Tables 1-4 display the clinical findings of these records. For the subjects with Parkinson's disease, the "severity" column in table is Hohn and Yahr score (a higher score indicates more advanced disease). For the subjects with Huntington's disease, this is the total functional capacity measure (a lower score indicates more advanced functional impairment). For the subjects with ALS, the number here is the time since the onset of the disease.

Table 1. Clinical findings for 10 patients with Parkinson's disease

| Subject ID | Age (YRS) | Height (meters) | Weight (Kg) | Gender | Gait Speed (m/s) | Severity |
|---|---|---|---|---|---|---|
| 1 | 77 | 2 | 86 | m | 0.95 | 4 |
| 2 | 44 | 1.67 | 54 | f | 1.26 | 1.5 |
| 3 | 80 | 1.81 | 77 | m | 0.98 | 2 |
| 4 | 74 | 1.72 | 43 | f | 0.91 | 3.5 |
| 5 | 75 | 1.92 | 91 | m | 1.05 | 2 |
| 6 | 53 | 2 | 86 | m | 1.33 | 2 |
| 7 | 64 | 1.67 | 54 | f | 0.91 | 4 |
| 8 | 64 | 1.83 | 73 | m | 0.84 | 4 |
| 9 | 68 | 1.92 | 84 | m | 1.05 | 1.5 |
| 10 | 60 | 1.94 | 74 | m | 1.19 | 3 |

Table 2. Clinical findings for 10 patients with Huntington's disease

| Subject ID | Age (YRS) | Height (meters) | Weight (Kg) | Gender | Gait Speed (m/s) | Severity |
|---|---|---|---|---|---|---|
| 1 | 41 | 1.86 | 72 | m | 1.68 | 8 |
| 2 | 42 | 1.78 | 58 | f | 1.05 | 11 |
| 3 | 66 | 1.75 | 63 | f | 1.05 | 4 |
| 4 | 47 | 1.88 | 64 | f | 1.4 | 2 |
| 5 | 36 | 2 | 85 | m | 1.82 | 10 |
| 6 | 41 | 1.83 | 59 | f | 1.54 | 8 |
| 7 | 71 | 2 | 75 | m | 1.05 | 2 |
| 8 | 53 | 1.81 | 56 | f | 1.26 | 9 |
| 9 | 54 | 1.8 | 90 | f | 1.26 | 12 |
| 10 | 47 | 1.78 | 102 | f | 1.05 | 4 |

Table 3. Clinical findings for 10 patients with ALS

| Subject ID | Age (YRS) | Height (meters) | Weight (Kg) | Gender | Gait Speed (m/s) | Duration (Month) |
|---|---|---|---|---|---|---|
| 1 | 68 | 1.80 | 86.18 | m | 1.302 | 1 |
| 2 | 63 | 1.83 | 83.92 | m | 1.219 | 14 |
| 3 | 70 | 1.57 | 40.82 | f | 0.853 | 13 |
| 4 | 70 | 1.70 | 58.97 | f | Missing | 54 |
| 5 | 36 | 1.70 | 74.39 | m | Missing | 5.5 |
| 6 | 43 | 1.75 | 68.95 | m | 0.77 | 17 |
| 7 | 65 | 1.73 | 81.65 | m | 1.302 | 9 |
| 8 | 51 | 1.83 | 106.6 | m | 1.085 | 3 |
| 9 | 50 | 1.58 | 61.24 | m | 0.899 | 54 |
| 10 | 40 | 1.7 | 61.24 | f | 1.219 | 14.5 |

Table 4. Clinical findings for 10 healthy control subjects

| Subject ID | Age (YRS) | Height (meters) | Weight (Kg) | Gender | Gait Speed (m/s) |
|---|---|---|---|---|---|
| 1 | 57 | 1.94 | 95 | f | 1.33 |
| 2 | 22 | 1.94 | 70 | m | 1.47 |
| 3 | 23 | 1.83 | 66 | f | 1.44 |
| 4 | 52 | 1.78 | 73 | f | 1.54 |
| 5 | 47 | 1.94 | 82 | f | 1.54 |
| 6 | 30 | 1.81 | 59 | f | 1.26 |
| 7 | 38 | 1.67 | 57 | f | 1.4 |
| 8 | 69 | 1.72 | 68 | f | 0.91 |
| 9 | 74 | 1.89 | 77 | m | 1.26 |
| 10 | 61 | 1.86 | 60 | f | 1.33 |

## 3. Estimation of Dimension

The first step of all time series analysis in this paper (and many others) is the so-called *embedding*. The term embedding means the process of constructing a set of vectors from the given time series. Embedding procedure needs establishing of at least two parameters –embedding dimension $m$ and time delay $\tau$- which may drastically differ from case to case [4]. Having chosen $m$ and $\tau$, embedding is defined as

$$x_n \rightarrow (x_n, x_{n-\tau}, x_{n-2\tau}, \ldots, x_{n-(m-1)\tau}),$$

where $x_n$ is the time series under consideration. After embedding, we can estimate the *correlation dimension* of embedded vectors (see [4] for different definitions of dimension and techniques used for calculating them). In this definition of dimension, the probability that two points in the data set are separated by a distance less than $\varepsilon$ is considered as dimension [4]. Define

$$C(\varepsilon) := lim_{N \rightarrow \infty} \frac{2}{N(N-1)} \sum_{i=1}^{N} \sum_{j=i+1}^{N} u(\varepsilon - \|x_i - x_j\|),$$

with $u$ the Heaviside step function, and $\mathbf{x}_i$ and $\mathbf{x}_j$ vectors in the embedded space (which contains $N$ vectors). Then, the correlation dimension is defined as follows

$$D_2 := lim_{\varepsilon \to 0} \frac{\log C(\varepsilon)}{\log \varepsilon}. \qquad (1)$$

The main reason for usefulness of the above definition is that the correlation dimension determines the number of independent variables necessary to describe the dynamics of the central nervous system [5].

A typical left-foot signal of Parkinson's disease is shown in Fig. 1 (a). Figure 1 (b) shows the embedded states of the time series shown in Fig. 1 (a) assuming $m = 4$ and $\tau = 1$. Clearly, these are feasible choices for dimension and delay since the data points in Fig. 1 (b) spread enough, and moreover, no redundancy or irrelevancy is observed in this figure.

All of the time series data introduced in Section 2 are embedded assuming $m = 4$ and $\tau = 1$, and then the corresponding correlation dimensions are estimated using (1). The results of these estimations are summarized in Table 5. As it is observed, the average dimension of Parkinson's disease and Huntington's disease is more than the average dimension of healthy control subjects while the average dimension of ALS is less than it.

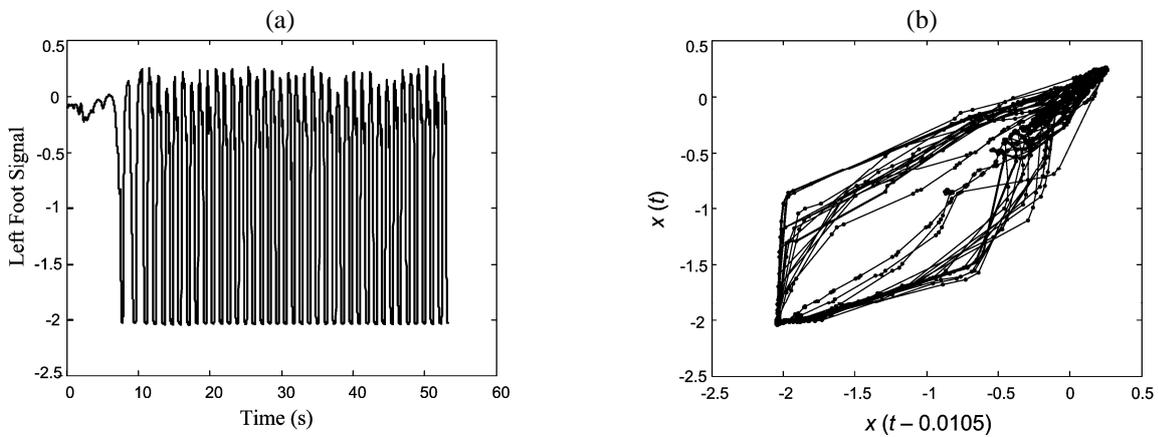

Fig. 1. (a) A typical left-foot signal of Parkinson's disease; (b) embedded states of the time series shown in Fig. 1 (a) assuming $m = 4$ and $\tau = 1$.

Table 5. Estimations of the correlation dimension

| Subject ID | Parkinson's disease | Huntington's disease | ALS | Healthy control subject |
|---|---|---|---|---|
| 1 | 2.76 | 3.13 | 2.08 | 1.78 |
| 2 | 2.88 | 3.41 | 1.78 | 1.74 |
| 3 | 2.76 | 3.13 | 2.08 | 1.78 |
| 4 | 1.72 | 2.77 | 1.54 | 1.82 |
| 5 | 1.74 | 2.52 | 1.89 | 1.82 |
| 6 | 2.06 | 2.31 | 1.79 | 3.04 |
| 7 | 1.90 | 2.21 | 1.72 | 2.62 |
| 8 | 2.51 | 2.49 | 2.42 | 1.94 |
| 9 | 2.33 | 2.68 | 2.56 | 2.04 |
| 10 | 2.70 | 2.09 | 2.03 | 2.41 |
| Average | **2.34** | **2.67** | **1.99** | **2.13** |
| S. T. D. | **0.45** | **0.44** | **0.32** | **0.45** |

## 4. Investigation of chaotic behavior by using 0-1 test

As it is observed in Fig. 1 (a) the signals under consideration are pseudo-periodic, which may seem to be generated from a chaotic dynamic. The 0-1 test is an effective tool for detecting the possible chaos in a given time series. One main advantage of the 0-1 test comparing to other tools (such as the Lyapunov exponents) is that it can be applied even if the equations governing the system are not known. As another advantage, this method can be applied to a given time series, say the output of a force-sensitive resistor, regardless of the method used for getting it.

A full exposition of the 0-1 test can be found in [6]. To explain briefly, denote the solution of the dynamical system by $\mathbf{x}(t)$ and consider $\phi(\mathbf{x})$ as the observable output of this system. As mentioned before, the method is essentially independent of the actual form of $\phi$ (i.e., almost any choice of $\phi$ will suffice). Now, define

$$\theta(t) = ct + \int_0^t \phi(x(s))ds,$$
$$p(t) = \int_0^t \phi(x(s))\cos(\theta(s))ds,$$

where $c > 0$ is an arbitrary chosen constant (we fix $c = 1.7$ in all simulations). To proceed, calculate the mean-square displacement of $p(t)$ as below

$$M(t) = \lim_{T \to \infty} \frac{1}{T} \int_0^T (p(t+\tau) - p(\tau))^2 d\tau,$$

and then plot $\log(M(t) + 1)$ versus $\log t$. The 0-1 test implies that the corresponding dynamical system is chaotic (non-chaotic) if the slope of the asymptotic line of this plot is equal to unity (zero).

The 0-1 test is applied to the time series data introduced in Section 2 and results are presented in Fig. 2 (a)-(d). The numbers on the subplots correspond to the subject IDs in Tables 1-4. According to Fig. 2 (a)-(d) it is obvious that all time series used in our simulations are non-chaotic since the slope of all asymptotes is equal to zero (note to the scales of horizontal and vertical axis).

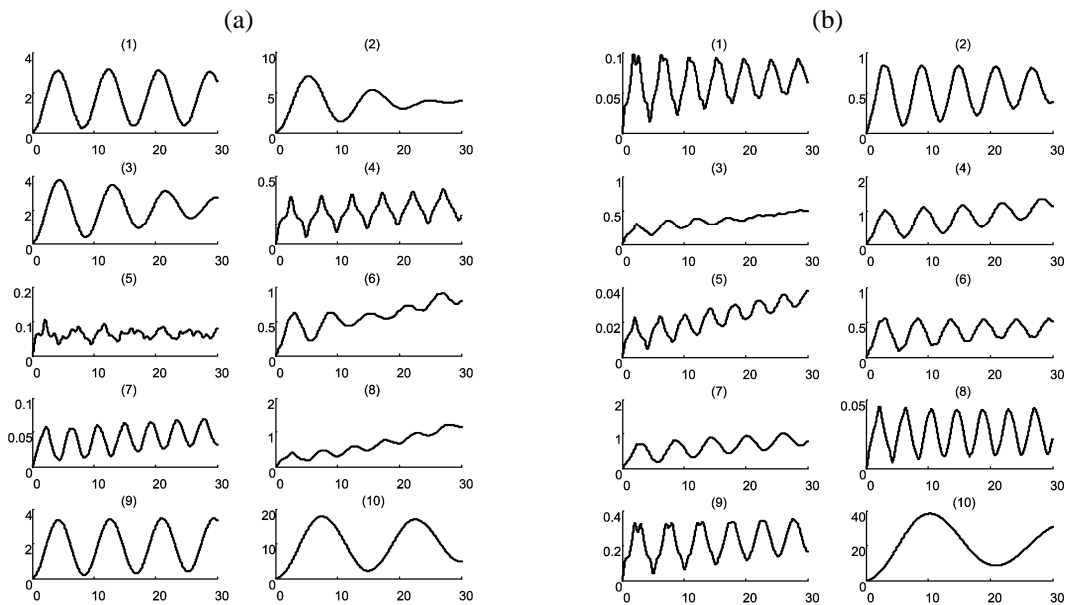

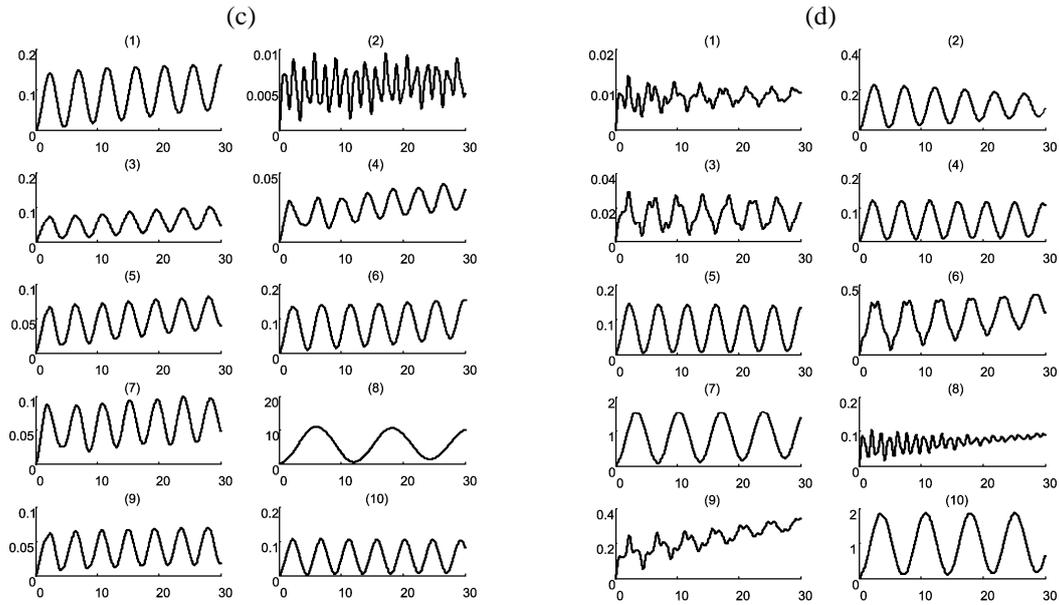

Fig. 2. The plot of $\log(M(t)+1)$ versus $\log t$, (a) Parkinson's disease; (b) Huntington's disease; (c) ALS; (d) healthy control subjects.

## 5. Conclusion

The time series data of three diseases with complex dynamics are studied is this paper. The correlation dimension of the time series are estimated and it is especially shown that the correlation dimension is less than four in all cases. The 0-1 test is also applied to the time series under consideration which concludes that none of them is chaotic.